\documentclass[12pt]{article}
\usepackage{epsf,amsmath}
\begin{document}

\title{Bell's theorem: Critique of proofs with and without inequalities}

\author{Karl Hess$^1$ and Walter Philipp$^2$}

\date{$^1$ Beckman Institute, Department of Electrical Engineering
and Department of Physics,University of Illinois, Urbana, Il 61801
\\ $^{2}$ Beckman Institute, Department of Statistics and Department of
Mathematics, University of Illinois,Urbana, Il 61801 \\ }
\maketitle

\begin{abstract}

Most of the standard proofs of the Bell theorem are based on the
Kolmogorov axioms of probability theory. We show that these proofs
contain mathematical steps that cannot be reconciled with the
Kolmogorov axioms. Specifically we demonstrate that these proofs
ignore the conclusion of a theorem of Vorob'ev on the consistency
of joint distributions. As a consequence Bell's theorem stated in
its full generality remains unproven, in particular, for extended
parameter spaces that are still objective local and that include
instrument parameters that are correlated by both time and
instrument settings. Although the Bell theorem correctly rules out
certain small classes of hidden variables, for these extended
parameter spaces the standard proofs come to a halt. The
Greenberger-Horne-Zeilinger (GHZ) approach is based on similar
fallacious arguments. For this case we are able to present an
objective local computer experiment that simulates the
experimental test of GHZ performed by Pan, Bouwmeester, Daniell,
Weinfurter and Zeilinger and that directly contradicts their claim
that Einstein-local elements of reality can neither explain the
results of quantum mechanical theory nor their experimental
results.

\end{abstract}

\section{Introduction}

Consider three joint pair probability distributions defined on the
Euclidean plane such that any two of these three joint pair
distributions share a common marginal. Let these three be
generated by pair distributions of the following pairs of random
variables $(A, B)$, $(A, C)$ and $(B, C)$. Then, according to a
theorem of Vorob'ev \cite{vorob} it may not be possible to realize
these three probability distributions on a common probability
space in the following sense. It may not be possible to find on
any probability space three random variables $A, B, C$ (and a
corresponding distribution formed for that triple) with the
property that the three joint pair distributions that can be
formed from that triple will coincide with the initially given
joint pair distributions for $(A, B)$, $(A, C)$ and $(B, C)$. Here
is a modification of the example that Vorob'ev used as the opening
statement for his paper. As before we label the three joint
distributions in terms of pairs of random variables (A,B), (A,C),
and (B,C). All random variables assume only the values +1 and -1.
We define these three joint probability distributions according to
the following table:

\begin{table}[ht]
    \begin{tabular}{|l||r|r|r|r|}\hline
   & $(+1,+1)$ & $(+1,-1)$ & $(-1,+1)$ & $(-1,-1)$
\\ \hline
      (A, B) & ${\frac {1} {4}}(1+{\frac {1} {\sqrt{2}}})$ & ${\frac {1} {4}}(1-{\frac {1} {\sqrt{2}}})$
       & ${\frac {1} {4}}(1-{\frac {1} {\sqrt{2}}})$ & ${\frac {1} {4}}(1+{\frac {1} {\sqrt{2}}})$\\ \hline
      (A, C) & ${\frac {1} {4}}(1+{\frac {1} {\sqrt{2}}})$ & ${\frac {1} {4}}(1-{\frac {1} {\sqrt{2}}})$
       & ${\frac {1} {4}}(1-{\frac {1} {\sqrt{2}}})$ & ${\frac {1} {4}}(1+{\frac {1} {\sqrt{2}}})$\\ \hline
      (B, C) & ${\frac {1} {4}}$ & ${\frac {1} {4}}$ & ${\frac {1} {4}}$ & ${\frac {1} {4}}$\\ \hline
   \end{tabular}
   \caption{Modified example of Vorob'ev \cite{vorob}.}\label{TA:ma}
\end{table}

Then it is easy to see that it is not possible to assign a
non-negative probability to the event $(A = -1, B= -1, C = -1)$
consistent with Table \ref{TA:ma}. Indeed, this latter probability
could not be more than ${\frac {1} {4}}$ because it cannot exceed
$P(B = -1, C = -1) = {\frac {1} {4}}$; similarly $P(A = -1, B =
-1, C = +1) \leq P(A = -1, C = +1) = {\frac {1} {4}}(1-{\frac {1}
{\sqrt{2}}})$. Adding these two probabilities we obtain $P(A = -1,
B = -1) \leq {\frac {1} {4}}(2-{\frac {1} {\sqrt{2}}})$ a bound
smaller than the value assigned in Table \ref{TA:ma}. Thus $A, B,
C$ {\it can not be defined on a common probability space} such
that the joint distribution of any of the three pairs that
possibly could be formed from them coincides with the pair
distributions defined by Table \ref{TA:ma}. Note that any two of
the three joint distributions share the same marginal, namely each
of the three variables $A, B, C$ assumes the values $+1$ and $-1$
with probability $1/2$. If, on the other hand, the twelve entries
in Table \ref{TA:ma} are all replaced by $1/4$, then three, even
independent, random variables $A, B, C$ {\it can be defined on
some common probability space} and each assuming the values $+1$
and $-1$ with probability $1/2$.

The following question arises immediately. If the twelve entries
in Table \ref{TA:ma} are replaced by non-negative numbers such
that the entries in each row add up to $1$, in other words if
Table \ref{TA:ma} constitutes a $3$x$4$ stochastic matrix, then
under which circumstances is it possible to realize the
corresponding joint distributions on a common probability space?
Bell, unknowingly, addressed this question in the context of
quantum mechanics by replacing the twelve entries by numbers that
depended on the covariances, resulting from such joint
distributions, and that were based on the negative cosines of
certain pairs of angles. He then assumed that a joint distribution
exists by defining $A, B, C$ as functions of a single random
variable $\Lambda$ as well as certain given settings as indicated
in Table \ref{TA:ma} with the goal of deducing consequences of
``some condition of locality, or of separability of distant
systems" \cite{bellbook}. Indeed, Bell tried to show then via his
inequality that objective local hidden variables such as $\Lambda$
can not exist \cite{bell} and states: ''In a theory in which
parameters are added to quantum mechanics to determine the results
of individual measurements, without changing the statistical
predictions, there must be a mechanism whereby the setting of one
measuring device can influence the reading of another instrument,
however remote". The basis for this far reaching statement was the
fact that for certain combinations of angles, depending on the
instrument settings, a contradiction between the predictions of
quantum mechanics and his inequality could be obtained. However,
as soon as $A, B, C$ are assumed to depend only on the single
random variable $\Lambda(\omega), \omega \in \Omega$, then
$A=A(\Lambda(\omega)), B=B(\Lambda(\omega))$ and
$C=C(\Lambda(\omega))$ are all defined on the same common
probability space $\Omega$. Thus in view of Vorob'ev's example the
contradiction is obtained even before we get started and without
recourse to any inequality. In the same vain, the derivation of
the Bell-type inequalities requires no further assumptions and is
independent of any additional considerations involving the
Einstein separation principle. In the present scenario, the basis
for Bell's proof is simply the assumption that $A, B, C$ can
simultaneously be measured in the sense that the values that these
three random variables assume can be simultaneously registered.
Therefore the definition of $A(\Lambda(\omega))$ etc. (denoted by
$A({\bf a}, \lambda)$ etc. in the first equation of Bell's
celebrated paper \cite{bell}) contains all the information needed
to derive the inequalities that contradict some quantum results.
However, according to Vorob'ev's example of Table \ref{TA:ma}
probabilities arising from certain ``closed loops" can not
consistently be described by random variables on a common
probability space. Hence the contradiction between the Bell
inequality and the predictions of quantum mechanics has its roots
entirely in purely mathematical reasons. At this point it is of no
concern whether or not $\Lambda$ depends on all or on none of the
instrument settings.

To better illustrate the ramifications of this discussion we ask
the reader to imagine the following situation. Assume that the
Aspect experiment \cite{eprex} had already been performed and
assume that Bell knew about it and also knew Vorob'ev's example of
Table \ref{TA:ma}. Bell wishes now to investigate the possibility
of a hidden variable model for the Aspect experiment. He knows
that in view of the Vorob'ev example the results of Aspect et al.
\cite{eprex} can not be explained by a model that uses three
random variables defined on a common probability space (see also
the discussion after Eq.(\ref{sauf2})). Therefore he rejects the
Ansatz $A=A(\Lambda(\omega)), B=B(\Lambda(\omega))$ and
$C=C(\Lambda(\omega))$. No inequalities are used or needed.

However, history proceeded along a different path. The Bell
inequality, as well as the more general CHSH \cite{chsh}
inequality, were obtained first and provided the decisive
motivation for the Aspect experiment \cite{eprex}. $\Lambda$'s
independence of the settings, a consequence of the delayed choice
of the settings in the Aspect experiment and of Einstein locality,
was considered crucial. Bell's Ansatz was considered to be most
general and the contradiction of Bell-type inequalities with the
data of the Aspect experiment was attributed by Bell to
non-localities.

We assume throughout as the basis for our analysis that the Aspect
experiment and all its results are valid and that no practical
deviations from the ideal embodiment of all experimental
procedures, such as detector inefficiencies etc., are of any
significance. We believe that ultimately it ought to be the goal
to find a physically reasonable mathematical model that can
explain the data of the Aspect experiment. We emphasize that our
criticism is directed only at some of the previous mathematical
models for the Aspect experiment, such as the Bell inequality, the
CHSH inequality and the arguments leading to these inequalities.

In many sciences it is a commonly accepted principle that if there
are competing theories that can be used to explain certain
phenomena then the simplest theory is the chosen one. We believe
that a simpler and thus a better explanation of the data of the
Aspect experiment can be  based on a model that in addition to a
source parameter $\Lambda$ includes time and setting dependent
instrument parameters $\Lambda_{\bf a}^*(t)$ and $\Lambda_{\bf
b}^{**}(t)$. These parameters are Einstein-local and they may be
quantum mechanical in nature, in the sense of describing atomistic
effects. $\Lambda_{\bf a}^*(t)$ and $\Lambda_{\bf b}^{**}(t)$ are
permitted to have their own distinctive stochastic behavior and
the same is true for all other settings.

We also note that independent of our considerations we believe
that the Aspect experiment is crucial for the interpretation of
quantum mechanics and particularly for the following important
distinction. It may either prove the existence of time and setting
dependent equipment parameters or, if their existence can be ruled
out for not yet known physical reasons, it may just show what the
followers of Bell have deduced all along e.g. non-locality or the
lack of validity of counterfactual reasoning as proposed by Peres
\cite{uperes}.

In contrast we shall show that other related experiments,
particularly those performed by Pan et al. \cite{pan} according to
the Greenberger-Horne-Zeilinger (GHZ) theory \cite{ghz} do permit
the construction of an objective local model. We shall point out a
serious flaw in their theory and, in addition, we will discuss a
simple experiment that can be performed on three independent
computers and that simulates the same experimental data as the
test performed by Pan et al. \cite{pan}. This computer simulation
directly contradicts their claim that their experiment establishes
non-locality without invoking inequalities. We also note that the
variations of Hardy, Peres and Mermin on the GHZ theory
\cite{peres1} suffer from similar deficiencies.

\section{Critique of proofs of the CHSH inequality}

\subsection{Probabilistic proofs}

This section is a brief summary of our previous work \cite{hpnp}
with the role of the Vorob'ev theorem \cite{vorob} woven in. First
we recall that the Bell inequality is contained in the CHSH
inequality as a special case. Therefore we focus our discussion on
the CHSH inequality noting that the same arguments apply mutatis
mutandis to the Bell inequality. From now on we use the standard
set-up and standard notation \cite{hpnp}. Correlated pairs of
particles are emitted from a source $S_0$, and the information
they carry is characterized by a random variable $\Lambda$.
Following Bell, we introduce random variables that describe spin
measurements $A = \pm 1$ in station $S_1$, and $B = \pm 1$ in
station $S_2$.  $A$ and $B$ are assumed to be functions of
$\Lambda$ and of the instrument settings that are denoted by
three-dimensional unit vectors, usually  $\bf a$ and $\bf d$  in
$S_1$, and  $\bf b$  and $\bf c$ in $S_2$. The instrument settings
have a special status in the sense that they are controlled by the
experimenters in $S_1$ and $S_2$ respectively. The experimenter in
$S_1$ chooses  $\bf a$ or $\bf d$ with probability $\frac {1} {2}$
and the experimenter in $S_2$, stochastically independent of the
choice in $S_1$, chooses $\bf b$ or $\bf c$ with probability
$\frac {1} {2}$.

The standard proofs of the CHSH inequality, as presented in most
text-books, proceed as follows. An entity
\begin{equation}
\Gamma:= A({\bf a},{\Lambda}(.)).B({\bf b},{\Lambda}(.)) + A({\bf
a},{\Lambda}(.)).B({\bf c},{\Lambda}(.)) + A({\bf
d},{\Lambda}(.)).B({\bf b},{\Lambda}(.)) - A({\bf
d},{\Lambda}(.)).B({\bf c},{\Lambda}(.)) \label{sau1}
\end{equation}
is defined. Since $A({\bf a},..)$ and $A({\bf d},..)$ can be
factored and since either  $B({\bf b},..) + B({\bf c},..) = 0$  or
$B({\bf b},..) - B({\bf c},..) = 0$, whereas the other one
accordingly equals $\pm 2$ it is then claimed that  $\Gamma = \pm
2$. Hence the absolute value $|{\Gamma}| = +2$, and so integration
of this equation with respect to a probability measure and an
application of an elementary inequality for integrals yields the
CHSH inequality,
\begin{equation}
|E(A({\bf a})B({\bf b})) + E(A({\bf a})B({\bf c})) + E(A({\bf
d})B({\bf b})) - E(A({\bf d})B({\bf c}))| \leq 2 \label{sau2}
\end{equation}
Here $E$ stands for the expectation value operator. This proof of
the CHSH inequality is based on a mathematical model that does not
adequately represent the Aspect experiment. We first argue on the
level of elementary calculus. In each run of the Aspect experiment
only one of the four products in $\Gamma$ can be measured, a fact
that is generally also appreciated \cite{peres}, \cite{leggett}.
$\Gamma$ itself is not measured directly since measurement of the
four products requires four incompatible arrangements. It follows
that, in particular, the various $A$'s and $B$'s need not be the
same, although they are denoted the same. Thus the conclusion that
necessarily $\Gamma = \pm 2$ is not correct.

If we try to interpret the above proof of the CHSH inequality as a
proof based on measure theoretic probability theory we first need
to agree on the underlying sample space. According to the
classical basic texts on probability theory by Feller
\cite{feller} and on measure theory by Halmos \cite{halmos} for
probability theory to apply to real world problems, a one-to-one
correspondence between the elements of the sample space and the
experiments to be performed must be established first. A random
variable is by definition a function (measurable in the sense of
measure theory) defined on that sample space, i.e. to each
performed experiment there is a well-defined value attached to it,
the value that the random variable assumes. Because, as was noted
above, the Aspect experiment does not measure $\Gamma$ itself,
without any additional assumptions on the stochastic relationship
between the potential hidden variables and the setting vectors,
$\Gamma$ may not be a well-defined function on any sample space,
thus may not be a random variable. This is not just a mathematical
technicality because the following issue is important. It has been
known (see e.g. the critiques by L. Accardi \cite{arc} and A. Fine
\cite{fine}), that the crucial ingredient in the standard proof of
the CHSH inequality, as reproduced above, was the assumption that
the four random variables $A({\bf a}), A({\bf d}), B({\bf b}),
B({\bf c})$ can be defined on the same probability space, thereby
equating the various  $A$  and $B$ in $\Gamma$. At first glance
this only seems to contradict the fact that the settings can not
be all simultaneously considered which is, in many cases, only a
technicality. For example, a fair die can be thrown six times or
six fair dice can be thrown once, the resulting statistics is all
the same. However, in order to establish that $\Gamma$ is a random
variable, that $\Gamma = \pm 2$ and in turn to take the
expectation value $E(\Gamma)$ resulting in the CHSH inequality
requires that the joint distributions of the four pairs $[A({\bf
a}),B({\bf b})], [A({\bf a}),B({\bf c})], [A({\bf d}),B({\bf b})]$
and $[A({\bf d}),B({\bf c})]$ can be realized as the marginal
distributions of the fourfold distribution of [A({\bf a}), A({\bf
d}), B({\bf b}), B({\bf c})]. For certain special cases this can
indeed be established (see proofs at the end of this section).
But, in general, this clearly contradicts the conclusion of the
Vorob'ev theorem \cite{vorob} because, picturesquely speaking, the
four joint distributions form a closed loop. To be more specific
we consider the following table.

\begin{table}[ht]
    \begin{tabular}{|l||r|r|r|r|}\hline
   & $(+1,+1)$ & $(+1,-1)$ & $(-1,+1)$ & $(-1,-1)$
\\ \hline
      (A({\bf a}), B({\bf b})) & $(1+\sigma_{{\bf a}{\bf b}})/4$
    & $(1-\sigma_{{\bf a}{\bf b}})/4$ & $(1-\sigma_{{\bf a}{\bf b}})/4$
    & $(1+\sigma_{{\bf a}{\bf b}})/4$\\ \hline
      (A({\bf a}), B({\bf c})) & $(1+\sigma_{{\bf a}{\bf c}})/4$
    & $(1-\sigma_{{\bf a}{\bf c}})/4$ & $(1-\sigma_{{\bf a}{\bf c}})/4$
    & $(1+\sigma_{{\bf a}{\bf c}})/4$\\ \hline
      (A({\bf d}), B({\bf b})) & $(1+\sigma_{{\bf d}{\bf b}})/4$
    & $(1-\sigma_{{\bf d}{\bf b}})/4$ & $(1-\sigma_{{\bf d}{\bf b}})/4$
    & $(1+\sigma_{{\bf d}{\bf b}})/4$\\ \hline
      (A({\bf d}), B({\bf c})) & $(1+\sigma_{{\bf d}{\bf c}})/4$
    & $(1-\sigma_{{\bf d}{\bf c}})/4$ & $(1-\sigma_{{\bf d}{\bf c}})/4$
    & $(1+\sigma_{{\bf d}{\bf c}})/4$\\ \hline
   \end{tabular}
   \caption{Illustration of the Vorob'ev theorem in terms of the covariances
    $\sigma$.}\label{TA:man}
\end{table}
Here accordingly in each row $\sigma$ equals the covariance of
$(A, B)$, depending on the settings as indicated in $(A(.),B(.))$.
Under the assumption that each $ A $ and each $ B$ assumes the
values $+1$ and $-1$ with probability $\frac {1} {2}$ the
covariances $\sigma$ uniquely determine the entries in Table
\ref{TA:man}. Quantum mechanics identifies the $\sigma$'s in each
of the four rows as the negative cosines of the angles between the
corresponding setting vectors. The Vorob'ev theorem in conjunction
with the Kolmogorov existence and consistency theorem states that,
in general, it is not possible to realize four such joint
distributions as marginals of a fourfold distribution of four
random variables defined on a single probability space. To prove
this claim directly we choose the $\sigma$'s in the first three
rows to equal $\frac {1} {\sqrt{2}}$ and the $\sigma$ in the last
row to be $-\frac {1} {\sqrt{2}}$. The above claim follows
immediately by an argument similar to the one given in the
introduction. Indeed from Table \ref{TA:man} we obtain that for
each of the four probabilities (corresponding to each choice of
$+1$ or $-1$)
\begin{equation}
P(A({\bf a}) = -1, B({\bf b}) = -1, B({\bf c}) = \pm 1, A({\bf d})
= \pm1) \leq (1 - \frac {1} {\sqrt{2}})/4 \label{sauf1}
\end{equation}
Adding these four probabilities we have
\begin{equation}
P(A({\bf a}) = -1, B({\bf b}) = -1) \leq 1 - \frac {1} {\sqrt{2}}
\label{sauf2}
\end{equation}
in contradiction to the value assigned by Table \ref{TA:man}.

Hence the fact that for some choices of angles between the setting
vectors the CHSH inequality is in contradiction with the
predictions of quantum mechanics is not a consequence of some
mysterious nonlocal physical phenomena, but rather a
straightforward consequence of basic mathematics.

Of course, in some particular cases it may be possible to realize
these four joint distributions as marginals of a fourfold
distribution, for instance if all four $\sigma$'s equal zero, i.e.
all 16 entries equal $\frac {1} {4}$, i.e. if the four random
variables are pairwise independent.

The following facts are now evident. In view of the Vorob'ev
theorem and example, {\it neither the Bell inequality nor the CHSH
inequality provide a conclusive tool to decide whether or not an
objective local model of any particular experiment can be
established}. What is really needed is a direct check of whether
or not the relevant random variables can be defined on the same
probability space. For the example of Table \ref{TA:ma} the Bell
inequality,
\begin{equation}
|E(AB) - E(AC)| \leq 1 - E(BC) \label{lsau}
\end{equation}
corresponding to the three angles $45^{\circ}, 45^{\circ},
90^{\circ}$ resulting in the covariances ${\frac {1} {\sqrt{2}}},
{\frac {1} {\sqrt{2}}}, 0$ is satisfied and yet, as we have seen,
$A, B, C$ can not be defined on the same probability space. Thus
for this case Bell's inequality is fulfilled yet Bell's Ansatz
still needs to be rejected.

For the mathematically inclined reader we offer the following
comments as to what is precisely going on here. Given three angles
or the corresponding unit vectors at least three Bell inequalities
can be written down. The one as displayed above and two more
obtained by cyclic permutation. Now it is not difficult to see
that if the direct approach shows that the three pair
distributions cannot be obtained from three random variables
defined on the same probability space then one of these Bell
inequalities will be violated. Hence the direct approach is
equivalent with checking all Bell inequalities, because as the
above example shows, checking only one of the inequalities may not
be enough to reach the desired conclusion.

On the other hand, for a very small class of parameters (e.g.
source parameters \cite{hpnp}) the conclusion of the Bell theorem
that is drawn from locality conditions remains valid. This can be
proved in several ways. In \cite{hpnp} we have presented proofs
based on what we called a reordering argument that work in some
special situations. For the benefit of the reader we will
reproduce this argument, based on elementary statistics, in the
following section. A second way to establish the Bell theorem in
this special situation is in essence the probabilistic counterpart
of \cite{hpnp} and is as follows. We assume that the hidden
variable consists only of a source parameter $\Lambda =
\Lambda(\omega)$ that is stochastically independent of the setting
vectors, considered as random variables $X(\omega^*)$ assuming one
of the two possibilities $\bf a$ or $\bf d$ in $S_1$, and
$Y(\omega^{**})$ assuming the values $\bf b$ or $\bf c$ in $S_2$,
with probability $\frac {1} {2}$ each. Because of the hypothesis
of the stochastic independence we can assume that there is a
common product probability space, say
$\Omega$x${\Omega^*}$x${\Omega^{**}}$ on which $\Lambda, X$ and
$Y$ are well-defined. Since $A$  and  $B$ are assumed to be
functions only of $x$ and $\lambda$ and only of $y$ and $\lambda$,
respectively, $A(X, \Lambda)$ and $B(Y, \Lambda)$ are random
variables defined on the same probability space. Hence applying
Fubini's theorem we conclude that $A({\bf a}, \Lambda(\omega)),
A({\bf d}, \Lambda(\omega)), B({\bf b}, \Lambda(\omega))$ and
$B({\bf c}, \Lambda(\omega))$ are also random variables that are
all defined on the same probability space. Thus in this special
case the entity $\Gamma$ turns out to be a random variable and
hence in this special case the proof of the CHSH inequality is
correct. A third way, and at the same time the most efficient one
to establish the Bell theorem in this special situation is based
on the above example that shows that the four joint pair
distributions in Table \ref{TA:man} cannot be obtained as the
marginals of four random variables $A({\bf a}), A({\bf d}), B({\bf
b})$ and $B({\bf c})$. Just above, in connection with Fubini's
theorem we remarked that substitution of $\Lambda$ would entail
the opposite statement. Hence such parameters $\Lambda$ can be
ruled out.

Another, more general instance where the Bell theorem remains
valid is when in addition to the source parameter $\Lambda$
instrument parameters $\Lambda^*$ in $S_1$ and $\Lambda^{**}$ in
$S_2$ are considered such that these three parameters are
stochastically independent. The same type of arguments will work
in this special case, too. Hence, these special classes of hidden
variables can be ruled out, that is for these classes the Bell
theorem is correct.

Of course, the validity of the Bell theorem in these special cases
does not imply that the Bell theorem is correct in general, that
is that all classes of hidden variables that can be reasonably
considered of being Einstein-local can be ruled out. For instance
the above type of arguments no longer work for what we call the
extended parameter space, that consists of a source parameter
$\Lambda$, and time and setting dependent instrument parameters
$\Lambda_{\bf a}^*(t)$ and $\Lambda_{\bf b}^{**}(t)$ in $S_1$ and
$S_2$, respectively. Because of the special role of time it is
admissible for the instrument parameters to be correlated by time
without violating the principle of Einstein-locality.

\subsection{Proofs based on elementary statistics}

Some standard text books consider the data accumulated by sampling
the above $\Gamma$. This is just the equivalent statistical
realization of the probability model considered in section 2.1.
However, in general, $\Gamma$ cannot be sampled because, in
general, $\Gamma$ is not a random variable. A reordering argument
works for the special situations considered above. As in section
2.1 let us assume that $\Lambda$ and the setting vectors $X$ and
$Y$ are stochastically independent and that $X$ assumes the vector
values $\bf a$ and $\bf d$ and $Y$ assumes the vector values $\bf
b$ and $\bf c$ with probability $1/2$ each. To avoid dealing with
$\epsilon$'s, let us assume in addition that $\Lambda$ assumes
only finitely many values $\lambda_s$ with positive probability
$p_s, s =1, 2,...,M$. If a large number $N$ of runs of the Aspect
experiment is performed then for each $s =1, 2,...,M$ we expect
the parameter value $\lambda_s$ to occur approximately $N.p_s$
times. Since by independence each pair of setting vectors ({\bf
a},{\bf b}), ({\bf a},{\bf c}), ({\bf d},{\bf b}) and ({\bf
d},{\bf c}) occurs about $\frac {1} {4}$ of the times each
$\lambda_s$ occurs, at the end of the day we can reorder and
rearrange the data points in approximately ${\frac {1} {4}} N.p_s$
rows all reporting to the same $\lambda_s$. Because now the
$\lambda_s$ are the same in each row the corresponding values
$\gamma$ are indeed $\pm 2$. Taking averages, denoted  by  $<…>$
and discarding rows that are possibly incomplete we obtain the
CHSH inequality in the form
\begin{equation}
|<a_j.b_j> + <a_j.c_j> + <d_j.b_j> - <d_j.c_j>| \leq 2
\label{sau3}
\end{equation}

Note that the lower case $a, b,...$ denote the outcomes
corresponding to the random variables with the corresponding
settings but not the settings themselves which are always
boldfaced. This is in agreement with the standard notation
\cite{peres}.

Again this line of argument fails for our extended parameter
space.

\subsection{Proofs based on sampling tables}

There are several arguments in the literature linking the validity
of the Bell theorem with sampling some kind of table and often
without any reference to the issue of hidden variables. In
contrast to the above situations in these arguments the emphasis
is placed solely on the set of potentially measured data. Most of
these arguments are deficient. Below we shall comment only on one
of these tables. But first, in analogy to Table \ref{TA:man}, we
describe the sampling procedure that mimics the Aspect experiment,
reviewing again the logical issues that need to be considered when
setting up a proper mathematical model. In the following table the
16 possible outcomes of the measurement pairs in the Aspect
experiment are represented as follows.

\begin{table}[ht]
  \begin{center}
    \begin{tabular}{|r|r|r|r|} \hline

       $a(+).b(+)$ & $a(+).b(-)$ & $a(-).b(+)$ & $a(-).b(-)$\\ \hline
       $a(+).c(+)$ & $a(+).c(-)$ & $a(-).c(+)$ & $a(-).c(-)$\\ \hline
       $d(+).b(+)$ & $d(+).b(-)$ & $d(-).b(+)$ & $d(-).b(-)$\\ \hline
       $d(+).c(+)$ & $d(+).c(-)$ & $d(-).c(+)$ & $d(-).c(-)$\\ \hline

   \end{tabular}
   \caption{16 possible outcomes of a Aspect experiment run}\label{TA:mas}
  \end{center}
\end{table}

Within each row a particular product is chosen with a certain
fixed probability such that the probabilities for each row add up
to 1. In other words the corresponding table of these
probabilities constitutes a  4x4 stochastic matrix. The example
given in  Table \ref{TA:man} represents the prediction by quantum
mechanics. Then each row in Table \ref{TA:mas} is sampled
according to the probability distribution (for instance, of Table
\ref{TA:man}) corresponding to that row and the average
corresponding to the outcomes of that row is calculated. Then the
average of the fourth row is subtracted from the sum of the
averages of the first three rows. Call the resulting quantity
$<\gamma>$. The CHSH inequality is equivalent to the statement
that no matter how the probabilities in the corresponding 4x4
stochastic matrix are chosen, the resulting $<\gamma>$ never
exceeds 2 in absolute value. However, it is easy to give examples
of 4x4 stochastic matrices with the property that if we sample
Table \ref{TA:mas} according to the probabilities of such a table
then $<\gamma>$ can be certainly bigger than 2, even as big as 4.
If, in addition, we also mimic the delayed choice provision of the
Aspect experiment then "after emission of a pair of particles" a
row of the above table gets chosen with probability $1/4$ and then
from this chosen row a sample of size 1 is taken according to the
corresponding probability distribution. After a series of such
"emissions" $<\gamma>$ is then calculated accordingly. The Aspect
experiment shows that the claim that $<\gamma>$ never exceeds 2 in
absolute value is false for some 4x4 stochastic matrices and the
Vorob'ev theorem provides the mathematical rationale for this
fact. Seen from this vantage point, the delayed choice provision
in the Aspect experiment is of no consequence.

Another demonstration of the conclusion of the Vorob'ev theorem is
given unwittingly in Table 6-1, on page 167 of the text-book by
Peres \cite{peres}. There Peres gives lower bounds on the
covariances of certain double sequences $(a_j,b_j), (b_j,c_j),
(c_j,d_j)$ and $(d_j,a_j)$ and then wonders why, at the end, the
covariance of the double sequence $(d_j,a_j)$ satisfies two
conflicting inequalities. Because, as observed above, the joint
distribution of a pair of random variables, each assuming with
probability $1/2$ the values +1 and -1 only, can be expressed in
terms of their covariance, and since the empirical distributions
of the four double sequences, again expressed picturesquely, form
a closed loop the Vorob'ev theorem says that, in general, it is
not possible to find a consistent joint distribution of four
random variables yielding the above imagined data set on which the
argument of Peres is based on.

\section{The GHZ Approach}

A decisive and penetrating analysis of the GHZ approach has been
given by Khrennikov \cite{khr}. Hence we shall review here only a
few of the basic ingredients needed below to describe the computer
experiment. In the paper "Bell's theorem without inequalities", by
Greenberger, Horne, Shimony, Zeilinger \cite{ghsz} the implicit
assumption that the $\lambda$ that occurs in all the relations
must be the same is clearly unfounded. The results of the actual
experiments are reported in \cite{pan}. For the photons $i =
1,2,3$ the authors introduce ``elements of reality $X_i$ with
values $\pm 1$ for $H'(V')$ polarizations and $Y_i$ with values
$\pm 1$ for R(L)" polarizations. The authors claim that the
elements of reality $X_i$ and $Y_i$ satisfy the relation
\begin{equation}
Y_1.Y_2.X_3 = -1 \text{,   } Y_1.X_2.Y_3 = -1 \text{,
}X_1.Y_2.Y_3 = -1 \label{sau4}
\end{equation}

Invoking counterfactual reasoning, the authors conclude that
\begin{equation}
{Y_i}.{Y_i} = +1 \label{saufn1}
\end{equation}
and thus by Eq.(\ref{sau4}) that
\begin{equation}
 X_1.X_2.X_3 = -1 \label{sau5}
\end{equation}

Counterfactual reasoning by itself is, in our opinion, not
objectionable. One certainly can argue that, if I had measured
with different settings and if I had the same photon(s) then I
would have obtained...In mathematical terms this just means that
for a given $\omega$, representing a given experiment, the random
variables $X_i$ and $Y_i$ will assume the values ${X_i}(\omega)$
and ${Y_i}(\omega)$, respectively, where $i = 1, 2, 3$. However,
this fact does not permit the conclusion that in Eq.(\ref{sau4})
the two $Y_1$ necessarily assume the same values, i.e.they are the
same random variables, because they are definitely obtained in two
distinctly different experiments. A similar statement holds for
$Y_2$ and $Y_3$. Thus subject to this interpretation
Eq.(\ref{saufn1}) is false as it stands.

Proceeding now with the discussion of the experiments of Pan et
al. \cite{pan} we note that Figure 3 of their paper depicts the
histograms for the actual $yyx, yxy$, and $xyy$ experiments. The
error rate for each of these three experiments is given with $0.15
\pm 0.02$. This translates into an error rate for the $xxx$ result
extracted from Eq.(\ref{sau4}) of  $0.45 \pm 0.035$ which is close
to $50\%$. Figure 4 of their paper shows the $xxx$ results
measured directly in a separate experiment. With success rate
$0.87 \pm 0.04$, it demonstrates that the product in
Eq.(\ref{sau5})  equals +1. This contradiction is statistically
highly significant and is the basis for their claim that, as a
consequence of their test, for the three-photon entanglement the
"quantum physical predictions are mutually contradictory with
expectations based on local realism."

We shall show now that their claim of non-locality is false by
providing an example that can be simulated on three independent
computers. Let
\begin{equation}
r_k(t) = sign[sin({2^k} \pi t)] \text{  for  }t>0 \label{sauf3}
\end{equation}
denote the $k$-th Rademacher function. Note that $r_k$ has period
$2^{-(k - 1)}$. The following table can serve as a basis for this
simulation

\begin{table}[ht]
  \begin{center}
    \begin{tabular}{|l||r|r|r|r|}\hline
   & $yyx,t_0<t<t_1$ & $yxy,t_2<t<t_3$ & $xyy,t_4<t<t_5$ & $xxx,t_6<t<t_7$
\\ \hline
      Comp1 & $Y_1=-r_1$ & $Y_1=-r_1$ & $X_1=r_2.r_3$ & $X_1=r_2.r_3$\\ \hline
      Comp2 & $Y_2=r_2$ & $X_2=r_1.r_3$ & $Y_2=r_2$ & $X_2=r_1.r_3$\\ \hline
      Comp3 & $X_3=r_1.r_2$ & $Y_3=r_3$ & $Y_3=-r_3$ & $X_3=r_1.r_2$\\ \hline
   \end{tabular}
   \caption{Computer simulation of the experiment of Pan et al. \cite{pan}}\label{TA:mal}
  \end{center}
\end{table}

Here  $t_1 - t_0$ is the length of time the $yyx$ experiment is
running, $t_2 - t_1$ is the length of time it takes the
experimenters to switch the experimental set-up from an $yyx$
experiment to an $yxy$ experiment. $t_3 - t_2$ is the length of
time the $yxy$ experiment is running and $t_4 - t_3$ again the
time to switch and so forth as described in Table \ref{TA:mal}.
Each of the three equations in Eq.(\ref{sau4}) holds on the entire
time interval where they are defined. Moreover, we have
\begin{equation}
 X_1.X_2.X_3 = +1 \label{saucorr5}
\end{equation}
instead of Eq.(\ref{sau5}), if we mimic at a later time the $xxx$
experiment according to the last column in Table \ref{TA:mal}.
Furthermore, each X and each Y equals +1 or -1 half of the time.
The essential point here is, of course, that for given equipment
settings, e.g. $yxy$, we can assume that equipment parameters are
such that $Y$ may be described by a certain Rademacher function,
e.g. $Y_3 = r_3$, while for the other $xyy$ we may have $Y_3 =
-r_3$. Here we have made use of the fact that in the actual
experiment the settings e.g. $yyx$ are set and used for a longer
period of time so that a mutual report can be established by
sub-light velocities between the measurement stations as to which
overall setting ($yxy$ or $xyy$ etc.) is used. For a given
setting, the outcomes of the various experiments are, of course,
only ``known" at a given detector, not at the others. Only the
choice of measurement time, which is random, determines the
outcomes together with the Rademacher functions that are
characteristic for a given setting. Of course the three Rademacher
functions in Table \ref{TA:mal} can be replaced by three other
Rademacher functions with arbitrarily large but different
subscripts if faster fluctuation between +1 and -1 is desired.

\section{Conclusions}

We have given a summary of the reasons why most of the proofs
leading to the Bell inequality and to the CHSH inequality are
deficient. In particular we have shown that in view of the theorem
of Vorob'ev the possibility of agreement of these inequalities
with the Aspect experiment is immediately lost as soon as $A, B$
are assumed to be the functions defined by Bell: his inequality
follows independently of any physics or locality conditions. We
have shown that time and setting dependent instrument parameters
that are Einstein-local {\it need not satisfy} Bell-type
inequalities. In fact, none of the known arguments leading to the
CHSH inequality can accommodate these parameters. Also, we have
presented a method more efficient than the Bell and the CHSH
inequality that can help to weed out specific classes of hidden
variables. This method does not rely on inequalities, but rather
on a simple determination whether a given set of joint pair
distributions can be realized as the marginals of the joint
distribution of random variables defined on the same probability
space. As mentioned in \cite{hpnp}, we do not have a proof that in
reality, not just mathematically, these setting and time dependent
parameters do exist for the Aspect experiment. Such a proof would
be established indirectly if, for instance, one would be able to
play the Bell game on two (stochastically and/or functionally)
independent computers with the same clock time. Given no further
information, we see this as a very difficult problem. However,
playing a Bell-type game for the GHZ approach is relatively easy
as we have demonstrated in the last section. For GHZ type of
experiments we do have an existence proof for setting and time
dependent instrument parameters because of the possibility of the
computer experiment. We believe that it is only a matter of time
that the same will be found for the Aspect experiment. Using the
(perhaps somewhat unreasonable) physical assumption of history
dependent instrument outcomes, we have found it already
\cite{hpp2}.

\section{ACKNOWLEDGEMENTS}

The authors would like to thank Andrei Khrennikov and Louis
Marchildon for valuable discussions. Support of the Office of
Naval Research (N00014-98-1-0604) is gratefully acknowledged.


\begin{thebibliography}{99}

\bibitem{vorob} N. N. Vorob'ev, {\it Theory of Probability and its
Applications} {\bf VII}, 147-162 (1962).

\bibitem{bellbook} J. S. Bell, {\it Speakable and Unspeakable in
Quantum Mechanics}, Cambridge University Press, Cambridge, U.K.,
1993, p 6.

\bibitem{bell} J. S. Bell, {\it Physics} {\bf 1}, 195-200 (1964).

\bibitem{eprex} A. Aspect, J. Dalibard and G. Roger,
{\it Phys. Rev. Letters} {\bf 49}, 1804-1807 (1982).

\bibitem{chsh} J. F. Clauser, M. A.  Horne, A.
Shimony and R. A. Holt {\it Phys. Rev. Letters} {\bf 23}, 880-883
(1969).

\bibitem{uperes} A. Peres {\it Am. J. Phys.} {\bf 46}, 745-747 (1978).


\bibitem{pan} Jian-Wei Pan, D Bouwmeester, M. Daniell,
H. Weinfurter, H. and A. Zeilinger, {\it Letters to Nature} {\bf
403}, 515-519 (2000).

\bibitem{ghz} D. M. Greenberger, M. A. Horne and A. Zeilinger,
in {\it Bell's Theorem, Quantum Theory, and Conceptions of the
Universe}, edited by M. Kafatos, Kluewer Academic Publishers,
Dordrecht, 1989, pp 73-76.

\bibitem{peres1} A. Peres, `{\it Quantum Theory: Concepts and Methods},
Kluewer Academic Publishers, Dordrecht, 1995, pp 151-153.

\bibitem{hpnp} K. Hess and W. Philipp, {\it Proceedings of the National Academy
of Sciences (USA)} {\bf 101} 1799-1805 (2004).

\bibitem{peres} A. Peres, `{\it Quantum Theory: Concepts and Methods},
Kluewer Academic Publishers, Dordrecht, 1995.

\bibitem{leggett} A. J. Leggett, {\it The Problems of Physics}, Oxford
University Press, 1987, pp 144-172.

\bibitem{feller} W. Feller, {\it An Introduction to Probability
Theory and its Applications} Vol 1, Wiley Series in Probability
and Mathematical Statistics, New York, 1968, pp 1-9.

\bibitem{halmos} P. R. Halmos, {\it Measure Theory}, Van Nostrand Co.,
Priceton, N. J., 1950, p 188.

\bibitem{arc} L. Accardi, {\it Phys. Rev.} {\bf 77}, 169-192
(1981).

\bibitem{fine} Fine, A. {\it J. Math. Phys.} {\bf 23}, 1306-1310 (1982).

\bibitem{khr} Andrei Khrennikov, {\it Physics Letters A} {\bf
278}, 2001, pp 307-314.

\bibitem{ghsz} D. M. Greenberger, M. Horne, M.A. Shimony and
A. Zeilinger, {\it Am. J. Phys} {\bf 58}, 1131 -1143 (1990).


\bibitem{hpp2} K. Hess and W. Philipp, {\it Proceedings of the National Academy
of Sciences (USA)}{\bf 98}, 14228-14233 (2001). See also: W.
Philipp and K. Hess, Proc. Conf. Foundations of Probability and
Physics -2, edited by A. Khrennikov, Vaxjo, Sweden, 2002, pp
469-488.



\end{thebibliography}
\end{document}